\documentclass[a4paper,12pt]{article}
\usepackage{amsthm}
\usepackage{amssymb}
\usepackage{amsmath}
\usepackage{amsfonts}

\newtheorem{remark}{Remark}[section]

\newtheorem{lemma}{Lemma}[section]
\newtheorem{theorem}{Theorem}[section]

\def\b1{\mbox{\boldmath $1$}}

\newenvironment{demo*}{\vspace{3mm}\noindent{\bf Proof.}}{\hfill $\Box$ \vspace{3mm}}

\begin{document}

\baselineskip=20pt
\title{\bf \Large Optimal dividends problem  with a terminal value for  spectrally positive L\'evy processes}
\author{\normalsize{\sc  Chuancun Yin}  and {\sc Yuzhen Wen }\\
[3mm] {\normalsize\it  School of Mathematical Sciences, Qufu Normal University}\\
{\normalsize\it Shandong 273165, P.R.\ China} \\
 e-mail: ccyin@mail.qfnu.edu.cn} \maketitle
 \centerline{\large {\bf Abstract}}
\vskip0.01cm
In this paper we consider a modified version of the classical   optimal  dividends problem
of de Finetti in which the  dividend payments subject to a penalty at ruin. We assume that
 the risk process is modeled by a general
spectrally positive  L\'evy process before dividends are deducted.  Using the fluctuation
theory of spectrally positive  L\'evy processes we give an explicit expression of the value function of a barrier strategy.
Subsequently we  show that  a barrier strategy is the optimal  strategy  among all admissible ones.
Our work is motivated by the recent work of     Bayraktar, Kyprianou and Yamazaki (2013). \\

\noindent {\bf Key Words}: \, {Barrier strategy, Dual model, Optimal dividend strategy,
 Scale functions,  Spectrally positive  L\'evy process,
Stochastic control.\\


\normalsize

\baselineskip=22pt

\section{\normalsize Introduction}\label{In}

In this paper  we consider a modified version of the classical   optimal  dividends problem
of de Finetti in which  the  dividend payments subject to a penalty at ruin.
Within this problem we assume that the underlying dynamic of the risk process is modeled by a spectrally positive L\'evy process.  In recent years, quite a few interesting papers deal with this type of model.  For example, Avanzi et al. (2007),  Avanzi and Gerber (2008),  Bayraktar and  Egami (2008), Li  and Wu (2009), Ng (2009), Yao,  Yang and Wang (2010),  Dai,  Liu  and  Luan (2010, 2011), Avanzi,   Shen and  Wong (2011),  Bayraktar, Kyprianou and Yamazaki (2013), Yin and Wen (2013) to name but a few.

 We now state the   optimal  dividends problem considered in this paper.
Let $X = \{X(t)\}_{t\ge 0}$ be a L\'evy process without negative jumps defined  on a
 filtered probability space $(\Omega,
{\cal F}, {\Bbb F}, P)$, where $\Bbb{F}=({\cal F}_t)$$_{t\ge 0}$ is
  generated by the process $X$ and  satisfies the usual conditions. As the process $X$ has no negative jumps, its Laplace exponent exists and
   is given by
\begin{equation}
  \Psi (\theta)=\frac1{t}\ln Ee^{-\theta X(t)} =c\theta + \frac1 2\sigma^2\theta^2
+\int_{0}^{\infty}(e^{-\theta x}-1+\theta x\text{\bf
1}_{\{|x|<1\}})\Pi (dx), \label{math-eq1}
\end{equation}
where ${\text{\bf 1}}_A$ is the indicator function of a set $A$,  $c>0$, $\sigma\ge 0$ and $\Pi$ is a measure on $(0,\infty)$ satisfying
$$\int_{0}^{\infty}(1 \wedge x^2)\Pi(dx)<\infty.$$
Denote by $P_x$ for the  law of $X$  when $X(0)=x$. Let $E_x$ be the
expectation  associated with   $P_x$. For short, we write $P$ and
$E$ when $X(0)=0$. To avoid trivialities, it is assumed that $X$  does not have monotone sample paths.
 In the sequel, we   assume that $-\Psi'(0+)=\Bbb{E}(X(1))>0$ which implies
the process $X$ drifts to $+\infty$. It is well known that if $\int_1^{\infty}y\Pi(dy)<\infty$, then $\Bbb{E}(X(1))<\infty$, and $\Bbb{E}(X(1))=-c+\int_1^{\infty}y\Pi(dy).$
Note that $X$ has paths of bounded variation if and only if
$$\sigma=0\;\;{\rm and}\;\; \int_{0}^{\infty}(1 \wedge x)\Pi(dx)<\infty.$$
In this case, we write (1.1)    as
\begin{equation}
  \Psi (\theta) =c_0\theta + \frac1 2\sigma^2\theta^2
+\int_{0}^{\infty}(e^{-\theta x}-1)\Pi (dx), \label{math-eq2}
\end{equation}
with $c_0=c+\int_0^1 x\Pi(dx)$ the so-called drift of $X$.
For more details on spectrally positive  L\'evy processes, the reader is referred to  Bertoin (1996), Sato (1999) and Kyprianou (2006).

The process $X$ is an appropriate  model  of a  company driven by inventions or discoveries, or the cash fund of an investment company before dividends are deduced.  Let $\pi=\{L_t^{\pi}:t\ge 0\}$ be a  dividend strategy  consisting  of a  nondecreasing,
 right-continuous and $\Bbb{F}$-adapted process starting at $0$, where $L_t^{\pi}$ standards for the lump-sums of dividends paid out by the company up until time $t$. The risk process with initial capital $x\ge 0$ and controlled by a dividend strategy $\pi$ is  defined by
  $U^{\pi}=\{U_t^{\pi}:t\ge 0\}$, where
 $$U_t^{\pi}=X(t)-L_t^{\pi},\ \ t\ge 0.$$
 The  ruin time is then given by
 $$\tau_{\pi}=\inf\{t>0|U_t^{\pi}=0\}.$$
 A dividend strategy is called admissible if $L_t^{\pi}-L_{t-}^{\pi}\le U_{t-}^{\pi}$, for all $t<\tau_{\pi}$,
 in other words the lump sum dividend payment
is smaller than the size of the available capital. We define the dividend-value function $V_{\pi}$ by
$$V_{\pi}(x)=E\left[\int_0^{\tau_{\pi}}e^{-q t}dL_t^{\pi}+Se^{-q \tau_{\pi}}|U_0^{\pi}=x\right],$$
 where $q>0$  is an interest force for the calculation of the present value and $S\in \Bbb{R}$ is the terminal value.
Let $\Xi$ be the set of all admissible dividend policies. De Finetti's dividend
problem consists of solving the following stochastic control problem:
\begin{equation}
V(x)=\sup_{\pi\in\Xi}V_{\pi}(x),
\end{equation}
and to find an optimal policy $\pi^*\in\Xi$ that satisfies $V(x)=V_{\pi^*}(x)$  for all $x\ge 0$.

Next, we shall have a review on the related literature.  This optimal dividend problem has recently gained a lot of attention in actuarial mathematics for spectrally negative L\'evy processes.   Avram et al (2007), Loeffen  (2008) and Kyprianou et al. (2010)
  studied the case of $S=0$  for spectrally negative L\'evy processes. The case $S<0$ was studied by Thonhauser and Albrecher (2007) for the compound Poisson model and Brownian motion risk process. The case $S\in\Bbb{R}$ was studied by Loeffen (2009, 2010) for spectrally negative L\'evy processes. It was shown that the optimal dividend strategy is formed by a barrier strategy for this type model under some  conditions imposed on the L\'evy measure. Moreover, Azcue and Muler (2005) have provided a counter-example for the case $S=0$ shows that a barrier strategy can  not be optimal. However, this in contrast with the dividend problem  in the case of $S=0$  for  spectrally positive L\'evy processes considered by
Bayraktar, Kyprianou and Yamazaki (2013), which shows that  there  a  barrier strategy always forms the optimal strategy, no matter the form of the jump measure. Motivated by the  work of     Bayraktar, Kyprianou and Yamazaki (2013),  the purpose of this paper is to examine the analogous
question  for the same model   in the case of $S\neq 0$.

The rest of the paper is organized as follows. In Section 2 we state some facts about scale functions.  In Section 3 we give the main results.
Explicit expressions for the expected discounted value of dividend payments are obtained,
and   it is shown that the optimal dividend
strategy is formed by a barrier strategy.

\setcounter{equation}{0}
\section{ Scale functions}\label{Sca}

 For an arbitrary spectrally positive L\'evy process,
the  Laplace exponent  $\Psi$ is  strictly convex and $\lim_{\theta\to\infty}\Psi(\theta)=\infty$. Moreover,  $\Psi$ is strictly increasing on $[\Phi(0),\infty)$, where
$\Phi(0)$ is the largest zero of $\Psi$.
Thus there exists a function $\Phi:[0,\infty)\rightarrow [\Phi(0),\infty)$
defined by $\Phi(q)=\sup\{\theta\ge 0: \Psi(\theta)=q\}$ (its right-inverse) and such that
$\Psi(\Phi(q))=q,\ q\ge 0.$

We now recall the definition of the $q-$scale function $W^{(q)}$ and some properties of this function. For  each $q\ge 0$ there exits a
continuous and increasing function $W^{(q)}:\Bbb{R}\rightarrow
[0,\infty)$, called the $q$-scale function defined in such a way
that $W^{(q)}(x) = 0$ for all $x < 0$ and on $[0,\infty)$ its
Laplace transform is given by
\begin{equation}
\int_0^{\infty}\text{e}^{-\theta
x}W^{(q)}(x)dx=\frac{1}{\Psi(\theta)-q},\; \theta
>\Phi(q). \label{math-eq4}
\end{equation}
Closely related to $W^{(q)}$ is the scale function $Z^{(q)}$ given by
$$Z^{(q)}(x)=1+q\int_0^x W^{(q)}(y)dy,\ x\in \Bbb{R}.$$
 We will frequently use the following functions
 $$\overline{W}^{(q)}(x)=\int_0^x W^{(q)}(z)dz,\  \ \overline{Z}^{(q)}(x)=\int_0^x Z^{(q)}(z)dz, \ x\in \Bbb{R}.$$
Note that
$$Z^{(q)}(x)=1,\ \ \overline{Z}^{(q)}(x)=x,\ \ x\le 0.$$
 If X has paths of bounded
variation then, for all $q\ge 0$, $W^{(q)}|_{(0,\infty)}\in
C^1(0,\infty)$ if and only if $\Pi$ has no atoms. In the case that
$X$ has paths of unbounded variation, then for all
$q\ge 0$, $W^{(q)}|_{(0,\infty)}\in C^1(0,\infty)$. Moreover if
$\sigma> 0$ then   $C^2(0,\infty)$.
Further, if the L\'evy measure has a density, then the scale
functions are always differentiable  (see e.g. Chan, Kyprianou and Savov (2011)).

The initial values of $W^{(q)}$ and its derivative   can be derived from (2.1):
\begin{equation}
 W^{(\delta)}(0+)=\left\{
  \begin{array}{ll}\frac{1}{c_0}, & {\rm if} \ X\ {\rm has\ paths\ of\ bounded\ variation},\\
    0, &{\rm otherwise},
  \end{array}
  \right. \nonumber
\end{equation}
and
 \begin{equation}
 {W^{(\delta)}}'(0+)=\left\{
  \begin{array}{lll}\frac{2}{\sigma^2}, & {\rm if} \ \sigma\neq 0,\\
    \frac{q+\Pi(0,\infty)}{c_0^2}, & {\rm if} \ X\ {\rm is \ compound\ Poisson}\\
    \infty, & {\rm if} \ \sigma=0\ {\rm and} \ \Pi(0,\infty)=\infty.
  \end{array}
  \right. \nonumber
\end{equation}
The functions $W^{(q)}(x)$ and $Z^{(q)}(x)$ play a key role in the solution of two-sided exit problem.
The following results can be extracted directly out of existing literature. See for example
 Korolyuk et al. (1976), Bertoin (1997),  Avram,  Kyprianou and Pistorius (2004),
 Kuznetsov, Kyprianou and Victor Rivero (2012) in a somewhat different form. Define the first passage times for $a<b$, with the convention $\inf
\emptyset=\infty$,
$$T^+_b=\inf\{t\ge 0: X(t)>b \},\;\;\;T_a^-=\inf\{t\ge 0: X(t)<a \},\ \ \tau_{ab}= T^-_a\wedge T_b^+.$$
Then we have for $x, y\in (a,b), q\ge 0,  z\ge b$,
\begin{eqnarray}&& E_x (e^{-q T^-_a}\text{\bf 1}_{\{T_a^-<T_b^+\}} )=\frac{W^{(q)}(b-x)}{W^{(q)}(b-a)},\\
 &&E_x (e^{-q T^+_b}\text{\bf 1}_{\{T_b^+<T_a^-\}})=Z^{(q)}(b-x)-W^{(q)}(b-x)\frac{Z^{(q)}(b-a)}{W^{(q)}(b-a)},\\
 &&E_x \left(e^{-q  \tau_{ab}}\text{\bf 1}_{\{X(\tau_{ab})=b\}} \right)=
 \frac{\sigma^2}{2}\left({W^{(q)}}'(b-x)- W^{(q)}(b-x)\frac{{W^{(q)}}'(b-a)}{W^{(q)}(b-a)}\right),\\
 &&E_x \left(e^{-q \tau_{ab}}\text{\bf 1}_{\{X(\tau_{ab}-)\in dy, X(\tau_{ab})\in dz\}}\right)=u^{(q)}(x,y)\Pi(dz-y)dy,
 \end{eqnarray}
 where
$$u^{(q)}(x,y)=W^{(q)}(b-x)\frac{W^{(q)}(y-a)}{W^{(q)}(b-a)}-W^{(q)}(y-x).$$
The identities (2.2) and (2.3) together with the strong Markov property imply  that
$$e^{-q (t\wedge\tau_{ab})}W^{(q)}(b-X(t\wedge\tau_{ab})), \ \ \  e^{-q (t\wedge\tau_{ab})}Z^{(q)}(b-X(t\wedge\tau_{ab}))$$
and
$$e^{-q (t\wedge\tau_{ab})}\left(Z^{(q)}(b- X(t\wedge\tau_{ab}))-W^{(q)}(b- X(t\wedge\tau_{ab}))\frac{Z^{(q)}(b-a)}{W^{(q)}(b-a)}\right)   $$
are martingales.

\setcounter{equation}{0}
\section{Main results }\label{Main}

Denoted by $\pi_b=\{L_t^b, t\le \tau^{b}\}$ the constant barrier strategy at level $b$ and let
$U_b=\{U_b(t):t\ge 0\}$ be the corresponding risk process, where
  $U_b(t)=X(t)-D_b(t)$, with $L_{0-}^b=0$ and $L_t^b=(b\vee\sup_{0\le s\le t}X(s))-b.$
  Note that $U_b(t)$ is a spectrally positive L\'evy process reflected at $b$,  $\pi_b\in\Xi$ and $L_0^b=x-b$ if $X(0)=x>b$.
   Denote by   $V_b(x)$ the expected discounted
value of dividend payments, that is,
$$V_b(x)=E_x\left[\int_0^{T_b}e^{-q t}dL_t^{b}+Se^{-q T_b}\right],\ 0\le x\le b,$$
 where $T_b=\inf\{t>0: U_b(t)=0\}$
with $T_b=\infty$ if $U_b(t)>0$ for all $t\ge 0$. Here  $q>0$
is the discount factor and $S\in \Bbb{R}$ is the terminal value.

Denote by $\Gamma$ the extended generator of the process $X$, which acts on $C^2$ function $g$ defined by
\begin{equation}
{\cal{A}} g(x)=\frac{1} {2}\sigma^2 g''(x)-c g'(x)
+\int_{0}^{\infty}[g(x+y)-g(x)-g'(x)y\text{\bf
1}_{\{|y|<1\}}]\Pi(dy).\label{Thr-eq1}
\end{equation}

\begin {theorem} \label{thrm3-1} Let $S=0$.  Assume that $V_b(x)$ is bounded and twice continuously
differentiable on $(0, b)$ with a bounded first derivative.
 Then $V_b(x)$ satisfies the following
integro-differential equation
\begin{equation}
{\cal A}V_b(x)=q V_b(x), \; 0<x<b,\nonumber
\end{equation}
together with the boundary conditions
$$V_b(0)=0, \ V_b'(b)=1,\ V_b(x)=x-b+V_b(b)\ {\rm for}\  x>b.$$
\end{theorem}
{\bf Proof}\ Similar to the case of jump-diffusion (cf. Yin,  Shen and   Wen (2013)),  applying It\^{o}'s formula for semimartingales one has
\begin{eqnarray*}
&&E_x\left[e^{-q(t\wedge T_b)}V_b(U_b(t\wedge T_b))\right]=V_b(x)\\
&&+E_x \int_0^{t\wedge T_b}e^{-q s} [({\cal A}-q)V_b(U_b(s))]ds- E_x\left[\int_0^{t\wedge T_b}e^{-q t}d L_t^{b}\right].
\end{eqnarray*}
Letting $t\to\infty$ and note that $V_b(0)=0$ we have
$$V_b(x)=E_x\left[\int_0^{T_b}e^{-q t}dL_t^{b}\right].$$
This ends the proof.

\begin {lemma} \label{thrm3-1} For $b, q\ge 0$ and $0\le x\le b$, we have
\begin{equation}
  E_x\left[e^{-q T_b}\right]=\frac{Z^{(q)}(b-x)}{Z^{(q)}(b)}.
\end{equation}

\end{lemma}
{\bf Proof}\ Let $Y_b(t)=b-U_b(t)$, then  $Y_b$ is a  reflected L\'evy process
 with initial value $b-x$. Define
 $\tilde{T}_b=\inf\{t>0: Y_b(t)\ge b\}$, then
 $$ E\left[e^{-q T_b}|U_b(0)=x\right]= E\left[e^{-q \tilde{T}_b}|Y_b(0)=b-x\right]=\frac{Z^{(q)}(b-x)}{Z^{(q)}(b)},$$
 where in the last step we have used the result of Proposition 2 (i) in Pistorius (2004), see also Theorem 2.8 (i) in Kuznetsov et al (2012).
This ends the proof.

The following result due to  Bayraktar, Kyprianou and Yamazaki (2013), here we give a different proof.

\begin {lemma}   For $b, q\ge 0$ and $0\le x\le b$, define
$$V(x,b)=E_x\left[\int_0^{T_b}e^{-q t}dL_t^{b}\right],$$ then
we have
\begin{equation}
 V(x,b)= \frac{\overline{Z}^{(q)}(b)}{Z^{(q)}(b)}Z^{(q)}(b-x)-\overline{Z}^{(q)}(b-x)
+\frac{\Psi'(0+)}{q}\left(\frac{Z^{(q)}(b-x)}{Z^{(q)}(b)}-1\right).
\end{equation}
\end{lemma}
{\bf Proof}\ By the law of total probability and the strong Markov property as in Yin et al (2013), we have
 \begin{equation}
 V(b,x)=h_1(x) V(b,b)+h_2(x),
 \end{equation}
where
$$h_1(x)=E_x \left(e^{-q T^+_b}\text{\bf 1}_{\{T_b^+<T_0^-\}}\right),$$
and
$$h_2(x)=E_x \left(e^{-q T^+_b}(X(T_b^+)-b)\text{\bf 1}_{\{T_b^+<T_0^-\}}\right).$$
By (2.3),
\begin{equation}
h_1(x)=Z^{(q)}(b-x)-W^{(q)}(b-x)\frac{Z^{(q)}(b)}{W^{(q)}(b)}.
\end{equation}
By (2.5),
\begin{eqnarray}
E_x \left(e^{-q T^+_b}X(T_b^+)\text{\bf 1}_{\{T_b^+<T_0^-\}}\right)&=&
\int_{y=0}^b\int_{z=b}^{\infty}zu^{(q)}(x,y)\Pi(dz-y)dy\nonumber\\
&&\equiv I_1(x)-I_2(x),
\end{eqnarray}
where
\begin{eqnarray}
I_1(x)&=&\int_{y=0}^b\int_{z=b}^{\infty}\frac{W^{(q)}(b-x)}{W^{(q)}(b)}W^{(q)}(y)z\Pi(dz-y)dy\nonumber\\
&=&-\frac{bZ^{(q)}(b)}{W^{(q)}(b)}W^{(q)}(b-x)+bcW^{(q)}(b-x)\nonumber\\
&&+\frac{W^{(q)}(b-x)}{W^{(q)}(b)}\left(\overline{Z}^{(q)}(b)-\frac{\Psi'(0+)}{q}Z^{(q)}(b)+\frac{\Psi'(0+)}{q}\right),
\end{eqnarray}
\begin{eqnarray}
I_2(x)&=&\int_{y=0}^b\int_{z=b}^{\infty}W^{(q)}(y-x)z\Pi(dz-y)dy\nonumber\\
&=&-bZ^{(q)}(b-x)+bcW^{(q)}(b-x)\nonumber\\
&&+\overline{Z}^{(q)}(b-x)-\frac{\Psi'(0+)}{q}Z^{(q)}(b-x)+\frac{\Psi'(0+)}{q},
\end{eqnarray}
 Substituting (3.7) and (3.8) into (3.6) we get
 \begin{eqnarray}
E_x \left(e^{-q T^+_b}X(T_b^+)\text{\bf 1}_{\{T_b^+<T_0^-\}}\right)&=&
 \frac{W^{(q)}(b-x)}{W^{(q)}(b)}\left(\overline{Z}^{(q)}(b)-\Psi'(0+)\overline{W}^{(q)}(b)-bZ^{(q)}(b)\right)\nonumber\\
 &&-\overline{Z}^{(q)}(b-x)+\Psi'(0+)\overline{W}^{(q)}(b-x)+bZ^{(q)}(b-x),\nonumber
 \end{eqnarray}
 from which and (3.5)  we arrive at
\begin{eqnarray}
h_2(x)&=&\frac{W^{(q)}(b-x)}{W^{(q)}(b)}\left(\overline{Z}^{(q)}(b)-\Psi'(0+)\overline{W}^{(q)}(b)\right)\nonumber\\
 &&-\overline{Z}^{(q)}(b-x)+\Psi'(0+)\overline{W}^{(q)}(b-x).
\end{eqnarray}

Substituting (3.5) and (3.9) into (3.4) and using the boundary condition $V'(b,b)=1$, we get
$$V(b,b)=\frac{\overline{Z}^{(q)}(b)}{Z^{(q)}(b)}+\frac{\Psi'(0+)}{qZ^{(q)}(b)}-\frac{\Psi'(0+)}{q},$$
and the result follows.

From Lemmas 3.1 and 3.2 we have
\begin {theorem} \label{thrm3-3} The expected discounted
value of dividend payments of the barrier strategy at level $b\ge 0$ is given by
\begin{equation}
 V_b(x)=\left\{
  \begin{array}{ll}\Lambda (b)Z^{(q)}(b-x)
  -\overline{Z}^{(q)}(b-x)-\frac{\Psi'(0+)}{q}, & {\rm if} \ 0\le x\le b,\\
    x-b+\Lambda (b)-\frac{\Psi'(0+)}{q}, &{\rm if} \ x>b,
  \end{array}
  \right.
\end{equation}
where
$$\Lambda (b)=\left(\overline{Z}^{(q)}(b) +\frac{\Psi'(0+)}{q}
  + S \right)\frac{1}{Z^{(q)}(b)}.$$
\end{theorem}
  By differentiating (3.10), we obtain that
  \begin{equation}
 V_b'(x)=\left\{
  \begin{array}{ll}-q\Lambda (b)W^{(q)}(b-x)
  +Z^{(q)}(b-x), & {\rm if} \ 0<x\le b,\\
    1, &{\rm if} \ x>b.
  \end{array}
  \right.
\end{equation}
  It follows that $V_b'(b)=1$ if and only if $\Lambda (b)=0$, or,
   equivalently $\overline{Z}^{(q)}(b)=-\frac{\Psi'(0+)}{q}-S.$
  We denote our candidate barrier level by
  \begin{equation}
 b^*=\left\{
  \begin{array}{ll} (\overline{Z}^{(q)})^{-1}(-\frac{\Psi'(0+)}{q}-S), & {\rm if} \ -\frac{\Psi'(0+)}{q}-S>0,\\
    0, &{\rm if} \ -\frac{\Psi'(0+)}{q}-S\le 0.
  \end{array}
  \right.
\end{equation}
  The expected discounted
value of dividend payments of the barrier strategy at level $b^*$ is given by
\begin{equation}
 V_{b^*}(x)=\left\{
  \begin{array}{ll} -\overline{Z}^{(q)}(b^*-x)-\frac{\Psi'(0+)}{q}, & {\rm if} \   -\frac{\Psi'(0+)}{q}-S>0,\\
    x+S, &{\rm if} \  -\frac{\Psi'(0+)}{q}-S\le 0,
  \end{array}
  \right.
\end{equation}
  for any $x\ge 0$.
\begin{remark} Letting $S\to 0$ in (3.12) and (3.13), the results deduced to (2.12) and (2.14) in  Bayraktar, Kyprianou and Yamazaki (2013).
\end{remark}

From the result Theorem 2.1 in  Bayraktar, Kyprianou and Yamazaki (2013), we have
\begin{theorem}
 Consider the stochastic control problem (1.3).    Then the barrier strategy at $b^*$ is an optimal  strategy for the control problem and
 $V(x)=V(x,b^*)$ as defined in (3.13).
\end{theorem}

\vskip0.3cm

\noindent{\bf Acknowledgements}\;
The research was supported by the National Natural
Science Foundation of China (No.11171179),  the Research Fund for
the Doctoral Program of Higher Education of China (No.
20093705110002) and the Program for  Scientific Research Innovation Team 
in Colleges and Universities of Shandong Province.

\end{document}